\documentclass[aps,prl,preprint,groupedaddress,superscriptaddress,showpacs,showkeywords]{revtex4}
\usepackage{graphics}
\usepackage{gensymb}
\usepackage{graphicx}
\usepackage{color}
\usepackage{amssymb}
\usepackage{fancyhdr}
\usepackage{layout}
\usepackage{graphicx}
\usepackage{subfigure}
\usepackage{amsmath}
\usepackage{indentfirst}
\usepackage{amssymb}
\usepackage{xcolor}
\usepackage[utf8]{inputenc}  
\usepackage[T1]{fontenc}
\usepackage{float}
\usepackage{textcomp}
\usepackage{times}

\begin{document}
\bibliographystyle{apsrev}

%\preprint{}

\title{Anomalous charge transport of superconducting Cu$_{x}$PdTe$_2$ under high pressure}


\author{Hancheng Yang}
\affiliation{IMPMC, Sorbonne Universit\'e and CNRS, 4 place Jussieu, 75005 Paris, France}
\author{M. K. Hooda}
\author{C. S. Yadav}
\affiliation{School of Basic Sciences, Indian Institute of Technology Mandi, Mandi-175005 (H.P.), India}
\author{David Hrabovsky}
\affiliation{Plateforme Mesures Physiques \`a Basses Temp\'eratures (MPBT), Sorbonne Universit\'e, 4 place Jussieu 75005 Paris, France}
\author{Andrea Gauzzi}
\author{Yannick Klein}
\email[]{yannick.klein@sorbonne-universite.fr}
\affiliation{IMPMC, Sorbonne Universit\'e and CNRS, 4 place Jussieu, 75005 Paris, France}

%\thanks{}

\date{\today}
\thispagestyle{empty}
\begin{abstract}
By means of high-pressure resistivity measurements on single crystals, we 
investigate the charge transport properties of Cu$_x$PdTe$_2$, notable for the combination of topological type-II Dirac semimetallic properties with superconductivity up to $T_c = 2.5$ K. In both cases of pristine ($x 
= 0$) and intercalated ($x=0.05$) samples, we find an unconventional $T^4$ power law behavior of the low-temperature resistivity visible up to 
$\sim$40 K and remarkably stable under pressure up to 8.2 GPa. This observation is explained by the low carrier density $n$, which strongly reduces the $k$-region available for electron-phonon scattering, as previously reported in other low-$n$ two-dimensional systems, such as multilayer graphene and semiconductor heterostructures. Our data analysis complemented by specific heat measurements and supported by previous 
quantum oscillation studies and \textit{ab initio} calculations suggests a scenario of one-band charge transport. Within this scenario, our analysis yields a large value of transport electron-phonon coupling constant $\lambda_{tr} = 1.2$ at ambient pressure that appears to be strongly enhanced by pressure assuming a constant effective mass.
\end{abstract}


\maketitle

\section{Introduction}
\label{introduction}
Layered transition metal dichalcogenides (TMD) $MX_2$ have attracted a great deal of interest for their rich physical properties, such as charge density waves (CDW) \cite{1CDW1,2CDW2,3CDW3}, superconductivity \cite{4super1,5super2}  and pressure-induced phase transitions \cite{6pressureinduced1,7pressureinduced2,8pressureinduced3}. Recently, the interest in these 
compounds has been renewed after Huang \textit{et al.}'s prediction \cite{12huang} of a novel type of topological electronic states, known as type-II Dirac cones. These states are characterized by a tilt of the energy dispersion curve of the cones with respect to the energy axis, which breaks the Lorentz invariance. These states display qualitatively different thermodynamic response as compared to that of type-I cones, typically found 
in graphene \cite{13graphene}. Namely, in type-II cones, the chiral anomaly depends on the direction of the magnetic field, which would lead to unusual topological transport properties.

Following the above seminal work, type-II Dirac cones have been predicted 
in a number of TMD's like PdTe$_2$, PtTe$_2$ and PtSe$_2$ \cite{9dirac_transport_PdTe2,10dirac_PtTe2,11PtSe2}, where the spin-orbit coupling (SOC) 
is sufficiently strong to produce the required tilt of the cones. Among these compounds, PdTe$_2$ has been intensively studied for it displays superconductivity below $T_c$=1.7-2 K \cite{9dirac_transport_PdTe2,14PdTe2Tc1,15dHvAPdTe2,16PTCPT}, attributed to a saddle-point-like van Hove singularity near the Fermi level, $E_F$ \cite{23vanhove}, enhanced up to 2.5 K upon Cu intercalation \cite{16PTCPT}. PdTe$_2$ has then been regarded as a promising playground to induce a topological superconducting (TSC) state at the surface by proximity \cite{18majorana}, though the experimental observation of this state remains controversial. On one hand, the existence of type-II Dirac cones have been predicted by \textit{ab initio} calculations and confirmed experimentally by angle resolved photoemission spectroscopy (ARPES) and by Shubnikov-de Haas (SdH) and de Haas-van Alphen (dHvA) quantum oscillations. The latter measurements further indicate that PdTe$_2$ is a multi-band semimetal with a nontrivial Berry phase for one of these bands \cite{9dirac_transport_PdTe2,15dHvAPdTe2,19QO_PdTe2}. On 
the other hand, scanning tunneling microscopy and spectroscopy studies \cite{20superPdte21,21superPdte22,22superPdTe23} support a picture of conventional BCS superconductivity contrasting the scenario of topological states at the Fermi surface.

In order to elucidate this controversy, a theoretical study suggests that 
high pressure may tune the superconducting and topological properties \cite{25calculHP}. Namely, the above study predicts a monotonic decrease of $T_c$ with pressure, the appearance of type-I cones above 4.7 GPa and the 
disappearance of the type-II cones above 6 GPa. These two abrupt changes in the electronic structure are expected to govern the interplay - if any 
- between superconductivity and topological Dirac states that can be probed experimentally. A first experimental study up to 2.5 GPa by Leng \textit{et al.} unveils a non-monotonic pressure dependence of $T_c$ displaying a maximum at $\sim$0.91 GPa \cite{26HP}, at odds with the above prediction. This discrepancy suggests that the calculations miss the details of the low-energy physics governing the superconducting state.

The purpose of the present study is to probe the signature of the evolution of the Dirac cones as a function of pressure in the electrical resistivity of pristine PdTe$_2$ and intercalated Cu$_{0.05}$PdTe$_2$ single crystals by means of a systematic study at much higher pressures up to 8.2 GPa. A favorable condition for a quantitative analysis of the results is that the transport properties of the system are dominated by one band, as previously reported \cite{15dHvAPdTe2}.

\section{Sample Synthesis Methods}
The PdTe$_2$ and Cu$_{0.05}$PdTe$_2$ single crystals object of the present study have been prepared following a two-step route. First, polycrystalline samples were synthesized via conventional solid-state reaction of a stoichiometric mixture of high-purity Pd, Te and Cu powders. The mixture was ground, pelletized and sealed in evacuated quartz tubes with residual 
pressure lower than $10^{-5}$ mbar. The pellets in the evacuated tubes were submitted to a heat treatment at 500 $^\circ$C for 3 days followed by a slow cooling-down to room temperature. In the second step, the as-prepared polycrystalline samples were reground and repelletized and submitted to a second heat treatment in evacuated quartz tubes at 757 $^\circ$C for 
one day, followed by a slow cooling-down to 500 $^\circ$C during one week 
and a final treatment at this temperature for another week. The heating was finally switched off to cool-down freely the samples to room temperature. The above synthesis route reproducibly yield 3-8 mm-sized platelet-shaped single crystals, as shown in Figs. 1(a,b). X-ray diffractograms obtained using a Rigaku X-ray diffractometer confirm the phase purity of the samples. Figs. 1(c,d) show the $c$-axis [00l] orientation of the platelets. For both compositions, PdTe$_2$ and Cu$_{0.05}$PdTe$_2$, the data analysis indicates that the crystal structure and the lattice parameters obtained are consistent with previous reports \cite{27PTCPT}. In Cu$_{0.05}$PdTe$_2$, we observed a minute shift of peak positions, attributed to a small reduction of the Pd-Te bond length \cite{16PTCPT}.
 
	\begin{figure}[ht] 
	\centering 
	\includegraphics[width = 0.7\columnwidth]{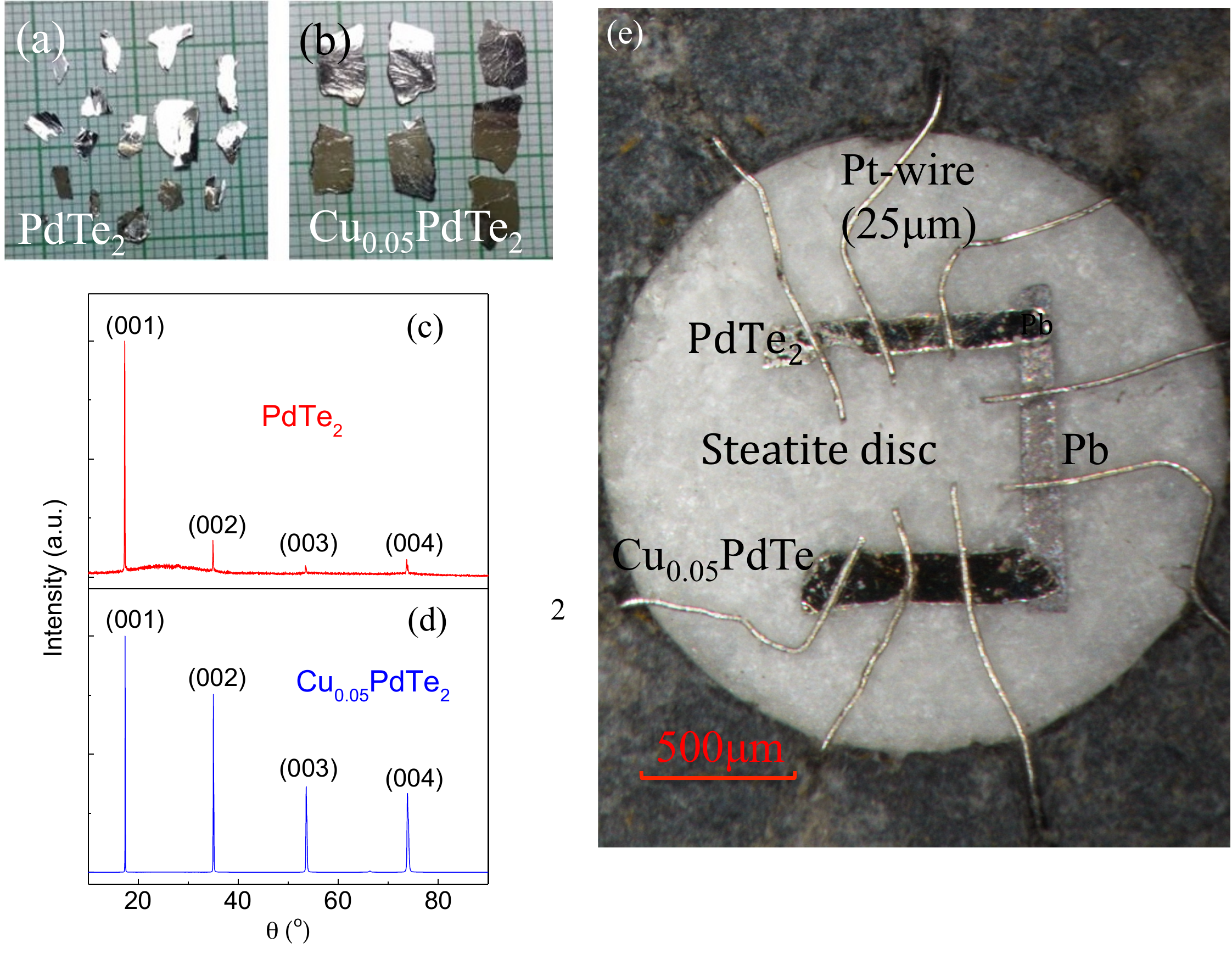}
	\caption{(a) and (b): Cleaved PdTe$_2$ and Cu$_{0.05}$PdTe$_2$ single crystals. (c) and (d): X-ray diffraction (XRD) pattern of PdTe$_2$ and Cu$_{0.05}$PdTe$_2$ showing $(00\ell)$ reflections. (e): Realization of the electrical contacts on the crystals in the Bridgman-type high-pressure cell (see text for details).}
	\label{fig1}
	\end{figure}
	
\section{Specific heat and high-pressure transport measurements}
For both measurements, we selected one single crystal from each of the PdTe$_2$ and Cu$_{0.05}$PdTe$_2$ batches, cleaved it and cut it in pieces of the desired shape and dimension. The isobaric specific heat, $C_p$ was measured in the 2-300 K range using a relaxation method implemented in a Quantum Design Physical Properties Measurement System (PPMS). The same apparatus was used for the electrical resistivity measurements using a conventional four-point method in the bar configuration. Typical bar dimensions are 700 $\times$ 300 $\times$ 20-40 $\mu$m$^3$. Owing to the platelet shape and orientation of the crystals, we measured the in-plane resistivity. For the ambient-pressure measurements, Au wires were attached on the cleaved surface using silver paste Dupont 6839. High-pressure measurements were carried out using a Bridgman-type cell, as described elsewhere \cite{bridgeman_cell}. The samples were positioned inside a pyrophyllite ring and sandwiched between steatite discs used as pressure transmitting medium (see Fig. 1e). In this case, the mechanical action of pressure alone ensures a good electrical contact between Pt wires and sample. Pressure was applied at room temperature by clamping the load using a locking nut and progressively increased up to 8.2 GPa after each measurement. The pressure value was determined by measuring the superconducting transition temperature $T_c$ of a thin Pb sample (Goodfellow, 99.99\% purity) placed near the PdTe$_2$ and Cu$_{0.05}$PdTe$_2$ crystals (see Fig. 1e) and using the pressure dependence of the $T_c$ of Pb as calibration curve.

\section{Results}
\subsection{Specific heat}
\begin{figure}[ht] 
\centering 
\includegraphics[width = 0.7\columnwidth]{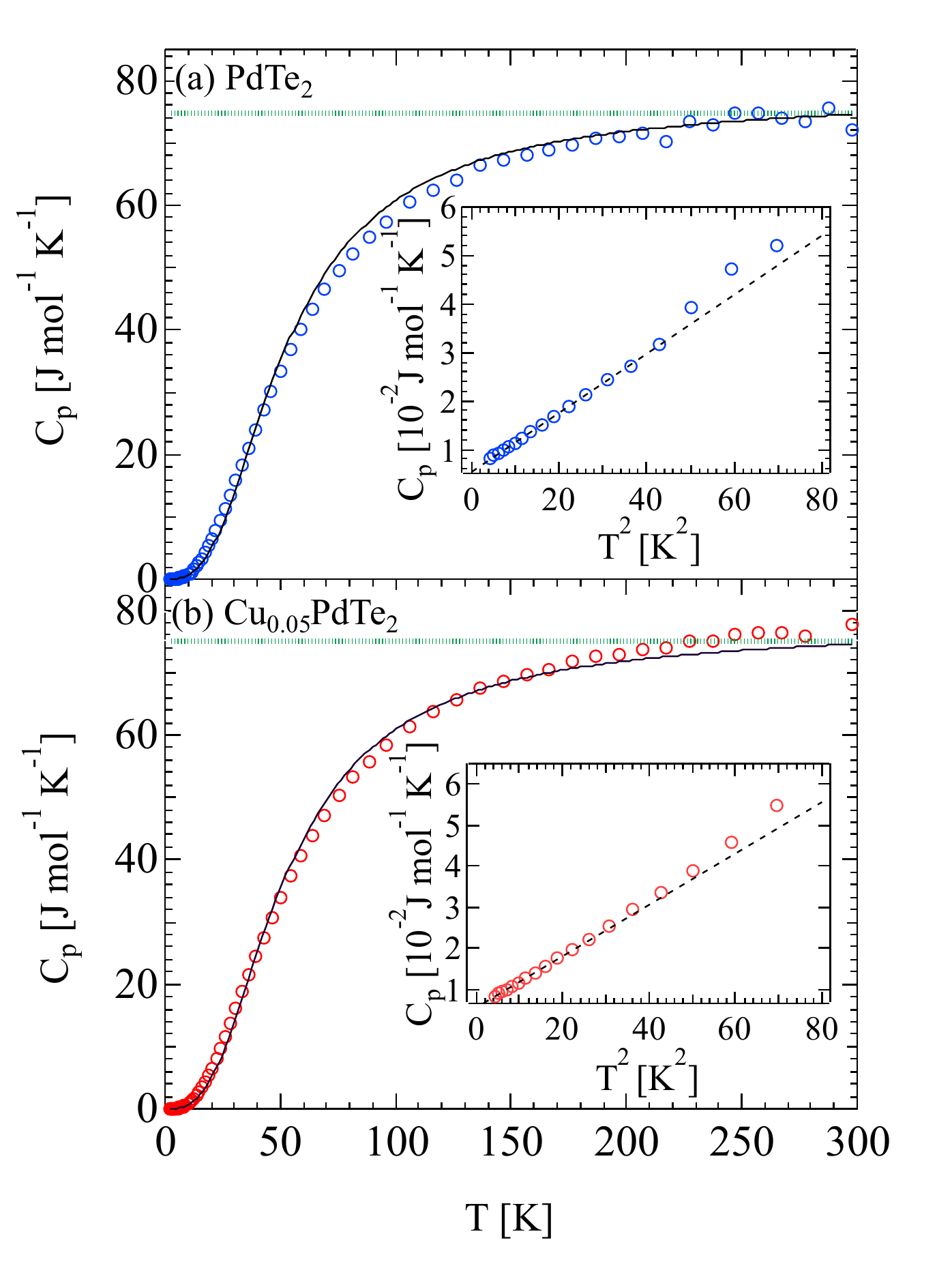}
\caption{Temperature dependence of the isobaric specific heat of PdTe$_2$ 
and Cu$_{0.05}$PdTe$_2$ (b) single crystals. The green broken lines and the solid lines indicate the Dulong-Petit limit and the prediction of the Debye model (see Eq.\ref{debye_cp} in the text), respectively. The insets 
show a detail of the $C_p/T\ vs\ T^2$ curves at low temperature. Black broken lines are linear fits to the expected $C_p/T = \gamma + \beta T^2$ 
behavior.}
\label{fig2}
\end{figure}

The $C_p(T)$ curves of both PdTe$_2$ and Cu$_{0.05}$PdTe$_2$ display a smooth behavior with no indication of phase transition in the whole 2-300 K 
range measured (see Fig. \ref{fig2}). In the high temperature region, the 
curves level off at the expected Dulong-Petit limit $3pR$ within the error made by neglecting the usually small difference between isobaric ($C_p$) and isochoric ($C_v$) specific heat. A straightforward analysis shows that, in the above temperature range, the data are well described by a conventional behavior resulting from the superposition of a linear electronic contribution and of a phonon contribution of the Debye type:

	\begin{equation}
	C_v=\gamma T+9pR \left(\frac{T}{\Theta_D} \right)^3 \int_{0}^{\frac{\Theta_D}{T}}\frac{x^4e^x}{(e^x-1)^2}dx
	\label{debye_cp}
	\end{equation}

where $\gamma$, $p$, $R$ and $\Theta_D$ are the Sommerfeld coefficient, the number of atoms per formula unit, the gas constant and the Debye temperature. The small difference between calculated and experimental points may be ascribed to the above difference between $C_p$ and $C_v$. $\gamma$ and $\Theta_D$ are determined by fitting the low-temperature data using the asymptotic dependence $C_v/T=\gamma + \beta T^2$. In good agreement with Ref.\cite{27PTCPT,29cppdte2}, the fit yields $\gamma=5.46(7)$ mJ mol$^{-1}$ K$^{-2}$ and $\beta=0.60(7)$ mJ mol$^{-1}$ K$^{-4}$ for PdTe$_2$ and $\gamma=5.74(3)$ mJ mol$^{-1}$ K$^{-2}$ and $\beta=0.62(2)$ mJ mol$^{-1}$ K$^{-4}$ for Cu$_{0.05}$PdTe$_2$. Using the Fermi liquid expression $\gamma=\pi^2 k_B^2 D(E_F)/3$, we obtain a density of states at 
the Fermi level of $D(E_F)= 2.31$ eV$^{-1}$ f.u.$^{-1}$ and 2.43 eV$^{-1}$ f.u.$^{-1}$  for PdTe$_2$ and Cu$_{0.05}$PdTe$_2$, respectively. The larger $D(E_F)$ in the latter compound is consistent with the electronic doping produced by Cu-intercalation. These values are consistent with early \textit{ab initio} calculations that include the SOC \cite{jan77}. Finally, using the Debye relation $\Theta_D = (12 \pi^4 pR/5 \beta)^{1/3}$, within the statistical uncertainty, we find the same value $\Theta_D=212-3(2)$ K for the two samples, as expected considering the modest concentration of intercalated Cu atoms.
	
\subsection{Resistivity}
\begin{figure}[ht] 
 	\centering 
 	\includegraphics[width = 0.7\columnwidth]{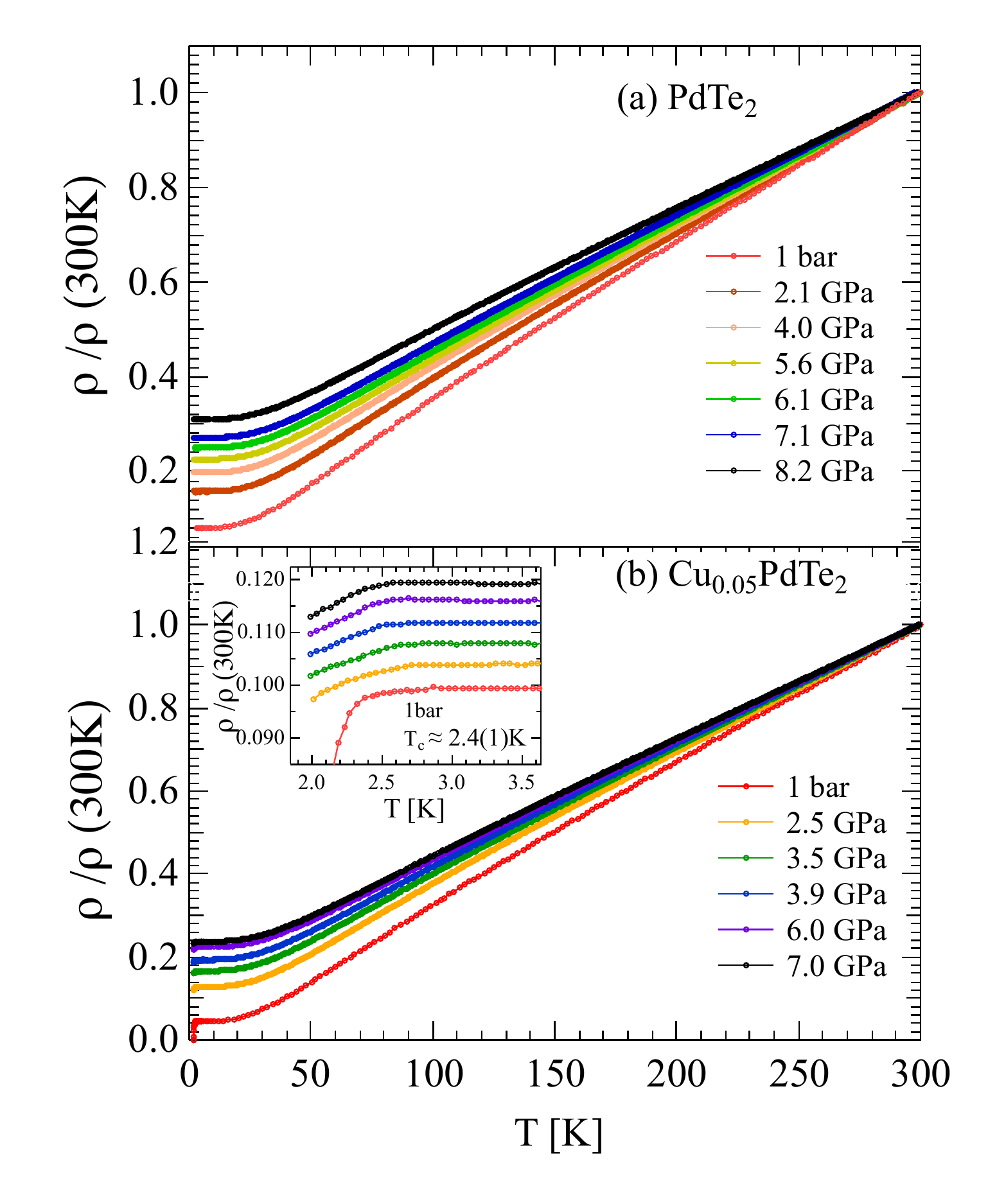}
 	\caption{Normalized electrical resistivity $\rho(T)/\rho(300 {\rm K})$ of PdTe$_2$ (a) and of Cu$_{0.05}$PdTe$_2$ (b) single crystals as a function of pressure. The inset in panel (b) shows a detailed view of the superconducting transition at low-temperature (curves are shifted vertically for clarity).}
 	\label{fig3}
 \end{figure}

In Fig. \ref{fig3} we plot the temperature dependence of the normalized in-plane resistivity $\rho(T)$ of the PdTe$_2$ and Cu$_{0.05}$PdTe$_2$ single crystals at different pressures. In PdTe$_2$, the ambient pressure and room temperature value of $44 \mu \Omega$ cm is consistent with the values in the $24 - 70 \mu \Omega$ cm range previously reported \cite{16PTCPT,29cppdte2,30RoPdTe2,31RoPdTe2}. The above scattering of values reported 
by different groups is attributed to the sensitivity of semimetals like PdTe$_2$ to slightly different concentrations of Te vacancies. The value $\rho_{300K} \approx 36 \mu \Omega$ cm measured in Cu$_{0.05}$PdTe$_2$ is explained by a significant carrier doping provided by Cu intercalation. For both compounds, the $\rho(T)$ curves evolve smoothly with pressure up to 8.2 GPa, which is expected considering that an earlier X-ray diffraction study gives no indication of any structural transition up to 19 GPa at 
room temperature \cite{32structure}. The absence of any anomaly with pressure indicates that the predicted crossover from type-II Dirac cones to type-I Dirac cones at $\sim 5$ GPa \cite{25calculHP} does not affect the transport properties or may occur at higher pressures. We finally note that the residual resistivity ratio ($RRR$) decreases with pressure, probably because of defects in the sample produced by a pressure gradient in the 
cell.

As seen in Fig. \ref{fig3}, no indication of superconductivity is found in PdTe$_2$ above 2 K in the whole pressure range measured, consistent with a theoretical prediction of even lower $T_c$'s $\sim$1.7-2 K at ambient pressure \cite{9dirac_transport_PdTe2,14PdTe2Tc1,15dHvAPdTe2,16PTCPT} and of a $T_c$ reduction with pressure \cite{25calculHP}. In the ambient pressure curve of Cu$_{0.05}$PdTe$_2$, we do observe a 
sharp superconducting transition with onset at $T_c \sim$ 2.4 K, in agreement with a previous report \cite{16PTCPT}. Under high-pressure, the transition broadens significantly so that zero resistivity is not achieved yet at 2 K. As above, we attribute this broadening to defects produced by a 
pressure gradient. The behavior of the $T_c$ onset value with pressure is 
seen in the inset of Fig. \ref{fig3}b. One may note a slight increase of $T_c$ at 2.5 GPa, followed by an equally slight decrease at higher pressures. Overall, these variations fall within a modest interval of 0.3 K, which corresponds to a rate smaller than 0.05 K/GPa.

\begin{figure}[ht] 
		\centering 
		\includegraphics[width = 0.7\columnwidth]{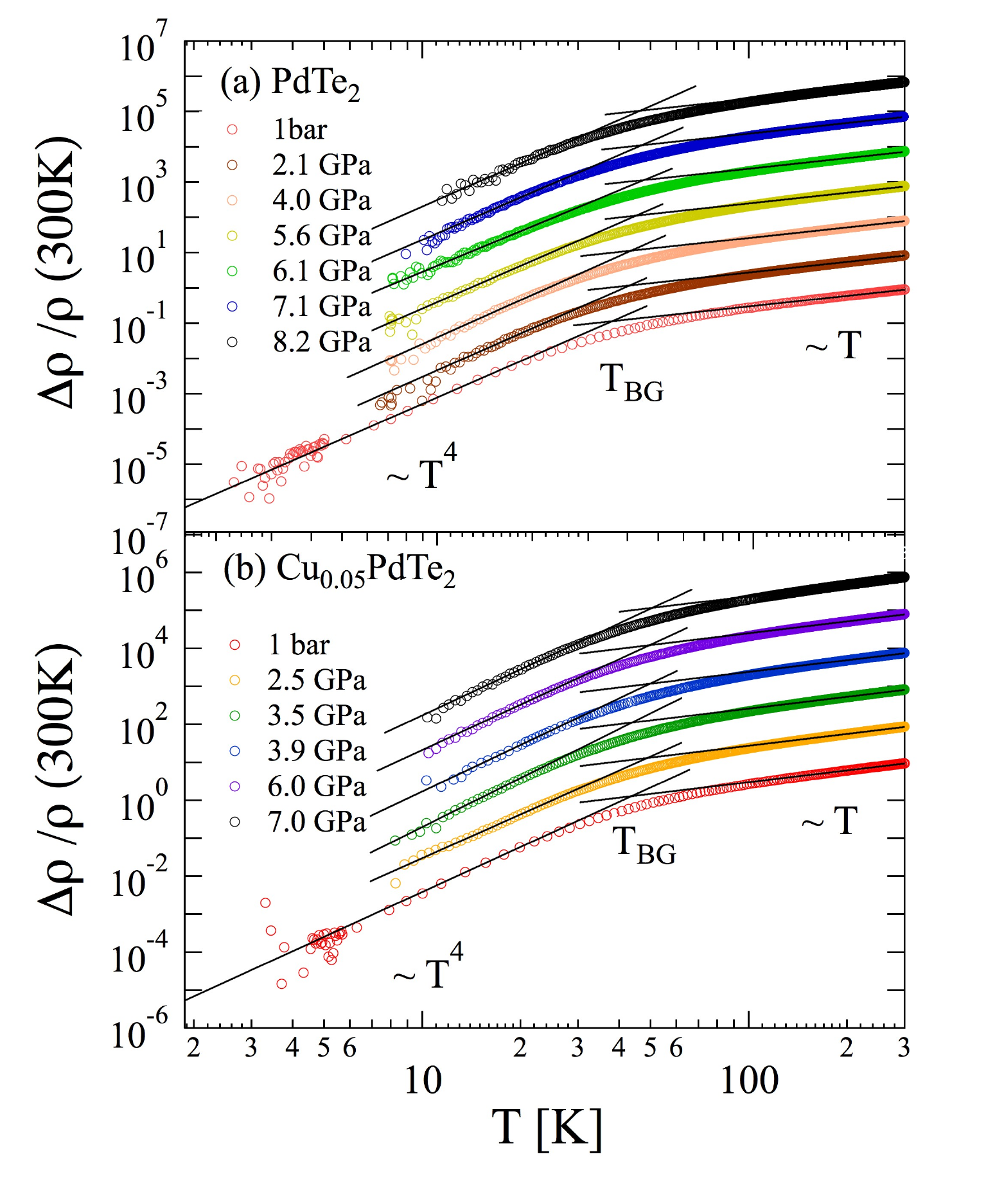}
		\caption{Log-log plot of the temperature dependent $\Delta \rho (T) = 
\rho (T) - \rho_0$ curves of PdTe$_2$ (a) and Cu$_{0.05}$PdTe$_2$ (b) at different pressures. Solid lines represent a fit to the linear and $T^4$ dependence. The crossover temperature between the two regimes define the Bloch-Gr\"uneisen temperature, $T_{BG}$. The curves are shifted vertically for clarity. The high-pressure data at low temperature are not shown due to the high noise.}
		\label{fig4}
\end{figure}

In Fig. \ref{fig3}, one further notes for both compounds 
a typical metallic behavior, characterized by a markedly linear dependence at high temperature followed by a saturation to a residual resistivity value $\rho_0$ at low temperature. As customary, we analyzed the behavior 
of this saturation using the power law $\rho(T) = \rho_0 + AT^n$. A straightforward data fit (see SM for details) shows that, in the whole pressure range investigated and for both compounds, the low-temperature behavior follows very well an unusual $T^4$ power law up to $\sim$40 K (see Fig. \ref{fig4}). At this temperature, the curves exhibit a smooth crossover to a linear dependence. This contrasts the $T^5$ power law expected in 
metals at low temperature within the Debye approximation. 

\section{Discussion}
To the best of our knowledge, the above unusual $T^4$ power-law has been reported previously in a few metallic systems and the proposed explanations are quite diverse \cite{33GdT4,34GdT4,35AgT4,36grapheneT4a,37graheneT4b,38grapheneT4c}. In polycrystalline Gd, a power law of the resistivity $\rho \sim T^n$ with $n=3.73 \pm 0.03$ observed in the 5 - 15 K range was explained by a linear spin-wave distribution law combined with a magnetic anisotropy \cite{33GdT4,34GdT4}. This scenario is clearly not applicable to the present case, for PdTe$_2$ and Cu$_{0.05}$PdTe$_2$ are both paramagnetic above $T_c$. A second example is given by Ag and Ag-based alloys that have been reported to display a $T^4$ power law similar to ours in 
the 2 - 6 K range \cite{35AgT4}. It was argued that this dependence limited to a narrow range of temperatures is ascribed to the concomitance of electron-phonon scattering and electron-electron scattering where Matthiessen's rule is not valid. This scenario is not applicable either to our case, where the observed $T^4$ dependence extends to a much wider temperature range, suggesting that a single scattering mechanism dominates.\\

The peculiar behavior observed here is rather similar to that found in multilayer graphene and bidimensional (2D) semiconductor heterojunctions with low carrier density, to which the so-called acoustic phonon limited model has been applied previously \cite{36grapheneT4a,37graheneT4b,38grapheneT4c}. All these systems share with PdTe$_2$ and Cu$_{0.05}$PdTe$_2$ a low carrier density and thus a small Fermi surface and a small Fermi wave vector $k_F$. When the Fermi surface is much smaller than the first Brillouin zone (BZ), only low-energy acoustic phonons participate in the electron-phonon scattering process. As a result, the low-temperature limit of the Debye model, corresponding to a $\Theta_D/T$ ratio approximated by infinity, is reached at temperatures much lower than the liquid helium temperatures of standard measurements like ours. In this case, instead of the 
Debye temperature $\Theta_D$, one has to introduce a cut-off Bloch-Gr\"uneisen temperature $T_{BG} =2 \hbar v_s k_F/k_B$, where $v_s$ is the sound velocity \cite{37graheneT4b}. A theoretical study proposed for graphene in the absence of screening shows that $\rho \propto T$ in the $T \gg T_{BG}$ limit, while $\rho \propto T^4$ in the opposite limit $T \ll T_{BG}$ \cite{36grapheneT4a}. This model, suitable to 2D semimetals in general, assumes that only the longitudinal acoustic mode participates in the electron-phonon scattering. Transverse acoustic modes are neglected because 
of the 2D characteristic of the system while optical modes are not populated at low temperatures \cite{36grapheneT4a}.

We argue that the above model is suitable indeed for the 
PdTe$_2$ system. First, in spite of the 3D character of the Fermi surface, a 2D model appears to be appropriate to describe the electrical transport because the electron-phonon coupling displays pronounced 2D properties, as shown by Kim et al. \cite{23vanhove}. Namely, by means of \textit{ab 
initio} calculations, these authors found that the frequency-dependent electron-phonon coupling constant $\lambda(\omega)$ is dominated by in-plane phonons. Second, we compare the value of $k_F$ with the size of the first BZ in the $ab$-plane. Considering that the lattice parameter is $a = 
4.037$ \AA, so $\pi /a = 0.78 \AA^{-1}$ \cite{32structure} and using the Onsager rule $F=A \hbar /2 \pi e$ relating the frequency $F$ of the quantum oscillations to the extremal Fermi surface area normal to the magnetic field, $A(E_F)$, we estimate $A(E_F)$ and, consequently, $k_F$. Assuming circular cross sections of the Fermi surface, SdH and dHvA measurements \cite{9dirac_transport_PdTe2,15dHvAPdTe2,19QO_PdTe2} unveil a low frequency at $\sim$ 8 T, a group of medium frequencies in the 110 T $< F <$ 140 T range and a high frequency of $\sim$450 T corresponding to $k_F$ values of 0.016, 0.057 - 0.062 and 0.12 \AA$^{-1}$, respectively. Evidently, these values are small as compared to the size of the first BZ, so the model of acoustic phonon-limited scattering is applicable. The SdH measurements also show very high frequencies of $\sim$1000 T and above, which corresponds to $k_F$ values comparable to $\pi/a$ \cite{19QO_PdTe2}. The charge carriers of these small pockets are then expected to dominate the transport properties in PdTe$_2$.

The crossover from $T^4$ to linear behavior in the log-log plot of the $\Delta \rho(T)$ curves in Fig \ref{fig4} indicates that $T_{BG} \sim 40$ K 
in PdTe$_2$ and we take $v_s \sim 3660$ m/s for the longitudinal acoustic 
branch, as reported previously \cite{39latticepdte2}. We then find $k_F \sim$ 0.068 \AA$^{-1}$, in good agreement with the magnitude of the medium 
frequencies of the above dHvA and SdH experiments. This suggests that the 
charge carriers of the corresponding extremal Fermi surface sections dominate the normal state transport properties. According to \textit{ab initio} calculations of the band structure and to the dHvA and SdH results, these sections belong to the six-fold plier-shaped Fermi pocket (see Fig.4(b) of \cite{9dirac_transport_PdTe2}), while the remaining frequencies are 
associated with other Fermi pockets \cite{9dirac_transport_PdTe2,19QO_PdTe2}. A further dHvA study shows that these Fermi pockets give large amplitudes in the fast Fourier transform (FFT) spectra \cite{15dHvAPdTe2}, in agreement with the Kohler plot of PdTe$_2$, which features a scaling of the magnetoresistivity in a wide range of temperature \cite{15dHvAPdTe2}, as commonly observed in single-band metals. The discrepancy between experimental and estimated $k_F$ values may be due to the following: (i) the contribution of charge carriers from other Fermi pockets. (ii) The Pd atoms are located in distorted octahedra, so the contribution of transverse acoustic modes with lower sound velocity may not be negligible. (iii) The above estimate of $k_F$ is based on the approximation of circular cross-sections of the Fermi pockets.

One further point of discussion concerns the effects of pressure and intercalation within the proposed scenario of acoustic phonon limited scattering. Although Hall measurements on PdTe$_2$ suggest a multiband picture \cite{9dirac_transport_PdTe2,26HP}, the above analysis suggests that a single band dominates charge transport, which allows us to estimate the electron-phonon coupling constant $\lambda$ using a simple one-band model for 
free-electron-like metals. Within such a model, at high temperatures ($T > \Theta_D$), the resistivity is expressed as $\rho =(2 \pi m^* k_B T/ne^2\hbar)\lambda_{tr}$ \cite{40EPC}, where $m^*$ and $n$ are the effective band mass and the charge carrier density, respectively. The transport electron-phonon coupling $\lambda_{tr}$ differs from the electron-phonon coupling $\lambda$ for the presence of the transport factor $1-\cos\theta$ in the scattering integral averaging over all phonon contributions, which gives extra weight to backscattering processes. Under 
ambient pressure and at high temperatures, $T > 200$ K, the $\rho(T)$ curve yields a resistivity coefficient $\partial \rho / \partial T = 0.14 \mu \Omega$ cm/K. So, taking $n=0.75 \times 10^{22}$ cm$^{-3}$ estimated at low field \cite{26HP} and $m^* = 0.3 m_e$ ($m_e$ is the free electron mass), which corresponds to the medium frequencies of the quantum oscillations \cite{9dirac_transport_PdTe2}, we obtain $\lambda_{tr} = 1.2$, to be compared with $\lambda = 0.53$ predicted by \textit{ab initio} calculations \cite{23vanhove} and $\lambda = 0.58$ reported in a very recent helium atom scattering study \cite{EP_PT}. We ascribe this discrepancy between $\lambda$ and $\lambda_{tr}$ values partly to the above transport factor and partly to the uncertainty in the geometry of the present resistivity measurements. In any case, the present estimate of $\lambda_{tr}$ gives a useful indication regarding the effect of pressure on the electron-phonon coupling. For instance, at 2 GPa, we find 
that $n$ increases up to $2.1 \times 10^{22}$ cm$^{-3}$ \cite{26HP}, and the slope of the $\rho$(T) curve at high temperature is 0.1 $\mu\Omega$ cm/K. Assuming that the $m^*$ remains constant, we conclude that $\lambda_{tr}$ increases by a factor 2 as compared to ambient pressure. This enhancement is at odds with the observed decrease of $T_c$ at 2 GPa \cite{26HP}. We can reconcile the two results by considering that the effective mass - and thus the band structure - is also expected to change with pressure.

\begin{figure}[ht] 
	\centering 
	\includegraphics[width = 0.7\columnwidth]{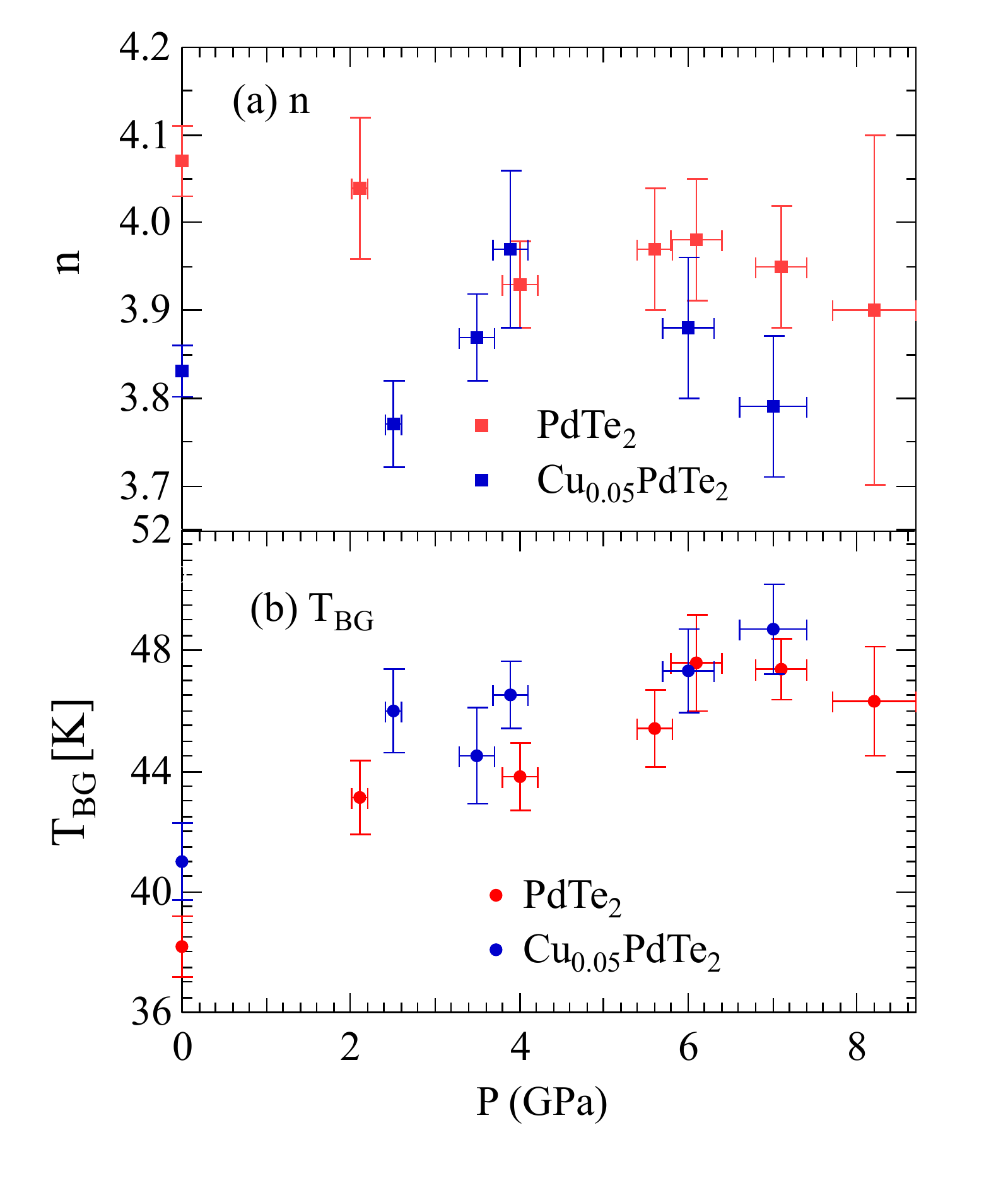}
	\caption{Pressure dependence of the power-law order coefficient $n$ (a) and of the Bloch-Gr\"uneisen temperature $T_{BG}$ (b) for PdTe$_2$ and Cu$_{0.05}$PdTe$_2$}.
	\label{fig5}
\end{figure}

We finally ask ourselves whether the 2D character of PdTe$_2$ is affected 
under high pressure or upon intercalation due to a larger orbital overlap 
along the $c$-axis and to the increasing importance of transverse modes. To test this possibility, in Fig. \ref{fig5}a we plot the power law coefficient $n$ as a function of pressure for PdTe$_2$ and Cu$_{0.05}$PdTe$_2$. Remarkably, $n$ is not sensitive to pressure and remains equal to 4 for both compounds within the statistical uncertainty. We argue that this robustness reflects the stability of the 2D character of the system. Indeed, it was reported \cite{32structure} that, at 8 GPa, the lattice constant ratio $c/a$ of PdTe$_2$ decreases by only 2.4\% as compared to the ambient pressure value. The $n$ value of Cu$_{0.05}$PdTe$_2$ is smaller than that for PdTe$_2$, which suggests that the 2D character is reduced by Cu intercalation, as expected. Fig. 5(b) shows the experimental values of $T_{BG}$ as a function of pressure for PdTe$_2$ and Cu$_{0.05}$PdTe$_2$. For both compounds, note an abrupt increase of $T_{BG}$ at low pressure, $\sim$2 GPa. Considering that the lattice parameters and 
the carrier density vary smoothly with pressure \cite{26HP,32structure}, one possible explanation is an anomalous variation of sound velocity at low pressure. Above 2 GPa, the moderate increase of $T_{BG}$ with pressure 
is ascribed to an increased carrier density. The value of $T_{BG}$ in Cu$_{0.05}$PdTe$_2$ slightly exceeds that of PdTe$_2$, which supports the picture that Cu intercalation enhances carrier density.

\section{Conclusion}
The present study on the type-II Dirac semimetals PdTe$_2$ and Cu$_{0.05}$PdTe$_2$ shows that the superconducting transition temperature $T_c$ is weakly affected by high pressure, with no anomaly in the pressure dependence of the resistivity curves up to 8.2 GPa. The latter result is at odds 
with the predicted pressure-induced evolution of $T_c$. Charge transport is also insensitive to the variations of the topological properties of the system under pressure, \textit{i.e.} the evolution of type II to type I 
Dirac cone, because the latter is located far from the Fermi energy. At low temperatures, all resistivity curves exhibit a remarkable $T^4$ power law up to 40 K and independent of pressure up to 8.2 GPa. Our data analysis supports a model of acoustic phonon-limited scattering suitable for low-density 2D materials, consistent with a scenario of charge transport dominated by one electronic band, as proposed previously \cite{15dHvAPdTe2}. Within this scenario, the dominant band is likely to be the six-fold plier-shaped Fermi surface pocket probed experimentally by quantum oscillations and predicted by \textit{ab initio} calculations. For this dominant band, our analysis yields a remarkably large value of transport electron-phonon coupling constant $\lambda_{tr} = 1.2$ at ambient pressure that appears to be strongly enhanced by pressure assuming a constant effective mass. Further studies are needed to confirm this enhancement while measurements at even higher pressures may definitely rule out the possibility that pressure can tune the topological properties of the system.

\section{Acknowledgement}
The authors are grateful to Stefan Klotz for useful discussions and Nicolas Dumesnil, Philippe Rosier and the staff of the low-temperature physical properties service (MPBT) of Sorbonne Universit\'e for technical support. The authors gratefully acknowledge the financial support provided by the Chinese Scholarship Council (Grant No. 201608070037) and by the Laboratory of Excellence MATISSE within the frame of the "Investissements d'Avenir Programme" of the French Ministery of University and Research (MESRI) 
under reference ANR-11-IDEX-0004-02.

\bibliography{reference}

\end{document}